\documentclass[11pt,floatfix,preprint,showpacs]{revtex4}
\usepackage{epsfig}
\usepackage{graphicx}
\usepackage{dcolumn}
\usepackage{bm}
\usepackage{amsmath}
\usepackage{amssymb}
\usepackage{amsfonts}

\begin{document}

\title{The exact three-dimensional half-shell t-matrix for a sharply
 cut-off
Coulomb potential in the
screening limit}

\author{W.\ Gl\"ockle$^1$}

\author{J.~Golak$^2$}

\author{R.~Skibi\'nski$^2$}

\author{H.~Wita{\l}a$^2$}

\affiliation{$^1$Institut f\"ur theoretische Physik II,
Ruhr-Universit\"at Bochum, D-44780 Bochum, Germany}

\affiliation{$^2$M. Smoluchowski Institute of Physics, Jagiellonian
University, PL-30059 Krak\'ow, Poland}

\date{\today}

\begin{abstract}
The three-dimensional half-shell t-matrix for a sharply cut-off Coulomb potential is
analytically derived together with its
 asymptotic form without reference to partial wave expansion.
The numerical solutions of the three-dimensional Lippmann-Schwinger
equation for increasing cut-off radii provide half-shell t-matrices
which are in quite a good agreement with the asymptotic values.
\end{abstract}

\pacs{21.45.+v, 24.70.+s, 25.10.+s, 25.40.Lw}

\maketitle \setcounter{page}{1}

\section{Introduction}
\label{intro}
This  is a continuation  of a previous article \cite{Gl09} where the
 exact
analytical three-di\-men\-sio\-nal
 wave function
for a sharply cut-off Coulomb potential has been derived together with
the
corresponding scattering
amplitude (the on-shell t-matrix). In the screening limit that
scattering
amplitude converges to a sum of
two terms. One is the expected pure Coulomb scattering amplitude
multiplied
with the standard
renormalisation factor $ ( 2 p R)^ { - 2i \eta}$; the other one is new
and
includes angular dependent
phases $ e^ { \pm 2i p R sin \frac{ \theta}{2}}$, which oscillate
without limit
for infinite screening
radius R. As has been  conjectured in \cite{Taylor} that second term
would
disappear after integration over
some angular intervals in the sense of a distribution.

We are now interested in the screening limit of the corresponding
  half-shell
t-matrix. The pure
Coulomb force result for that object is well known \cite{Kok}. Its
derivation
goes back to work by
\cite{Guth}.
Especially its discontinuous property at the on-shell  point is of
interest.
In \cite{Ford64,Ford66}
 this property
 has been discussed based on a sharply cut-off Coulomb potential and
 using  a
partial
 wave decomposition. Like in our previous paper we felt that a
direct three-dimensional approach avoids possibly open questions in that
treatment related to
 the correct summation of the infinite
number of partial wave components. (See \cite{Ford64,Ford66}, where
the
difficulties are spoken out
leading in fact to incomplete results).  We therefore study
the half-shell t-matrix now based on the exact three-dimensional wave
function for a sharply cut-off Coulomb potential and
investigate its screening limit. The details of derivation are given
in Section~\ref{hshtmatrix}. Numerical solutions of the three-dimensional
Lippmann-Schwinger equation for different cut-off radii are compared
with the asymptotic values in Section~\ref{numresults}. We summarize
and conclude in
Section~\ref{summary}.

\section{The half-shell t-matrix}
\label{hshtmatrix}

For a a sharply cut-off repulsive Coulomb potential (for instance
  for
two protons)
\begin{eqnarray}
V_R(r) = \Theta( R-r) \frac{e^ 2}{r}
\label{eq1}
\end{eqnarray}
the exact three-dimensional wave function inside the potential range is given by
\begin{eqnarray}
\Psi_R^ { (+)}  =   A e^ { i \vec p \cdot \vec r}
F( - i \eta,1,i ( pr - \vec p \cdot \vec  r))
\label{eq2}
\end{eqnarray}
with
\begin{eqnarray}
A = \frac{1}{ ( 2 \pi)^ { \frac{3}{2}}} \frac{1}{ F( - i \eta,1,2ipR)}
~.
\label{eq3}
\end{eqnarray}
The normalisation corresponds to the choice of
 $ \frac{1}{ ( 2 \pi)^ { \frac{3}{2}}}  e^ { i \vec p \cdot \vec r}$
as
incoming wave.
 Further $ \eta = \frac{ m e^ 2}{ 2p}$ for two particle with mass m. Then the
half-shell t-matrix is defined as
\begin{eqnarray}
< \vec p~' | V_R | \Psi_R^ { (+)}> = \frac{ 1}{ ( 2 \pi)^ { 3/2}} A
 \int d^ 3 r e^ {- i \vec p' \cdot \vec r} V_R(r) e^ { i \vec p \cdot \vec r}
 F( - i \eta,1,i ( pr - \vec p \cdot \vec  r)) ~.
\label{eq4}
\end{eqnarray}

We use the integral representation for the confluent hypergeometric function
\begin{eqnarray}
F( - i \eta,1,i ( pr - \vec p \cdot \vec  r)) = C( - i \eta,1)
 \int_{ \Gamma} dt ( \frac{1-t}{t})^ { i \eta} \frac{1}{t}
e^ { i ( pr - \vec p \cdot \vec r) t}
\label{eq5}
\end{eqnarray}
with
\begin{eqnarray}
C( - i \eta,1) = \frac{ -i}{ 2 \pi} e^ { \pi \eta}
\label{eq6}
\end{eqnarray}
and $ \Gamma$ a closed path in the complex $t$-plane encircling $t=0$
and $t=1$ in the positive sense.

The $r$-integral is straightforward leading to
\begin{eqnarray}
 \int^ R d^ 3 r e^ { i ( \vec p - \vec p')  \cdot \vec r} \frac{1}{r}
 e^ { i  prt} e^ { -i  \vec p \cdot \vec r t}
 =   \frac{ - 4 \pi}{ 2 \Omega} [ \frac{ e^ {i ( pt + \Omega) R} - 1}{
     pt
+ \Omega}
 - \frac{e^ { i ( pt - \Omega) R} - 1}{ pt - \Omega}]
\label{eq7}
\end{eqnarray}
with
 \begin{eqnarray}
\Omega &  = &  \sqrt{ p^ 2 t^ 2 - 2 t \vec p \cdot \vec \Delta +
  \Delta^ 2} ~, \\
\vec \Delta & = &  \vec p - \vec p~' ~.
\end{eqnarray}

Thus
\begin{eqnarray}
< \vec p~' | V_R | \Psi_R^ { (+)}> \equiv -
2 \pi \frac{ e^ 2}{ ( 2 \pi)^ { 3/2}}AC( - i \eta,1) Y
\label{eq10}
\end{eqnarray}
and
\begin{eqnarray}
Y  =   \int_{ \Gamma} dt ( \frac{1-t}{t})^ { i \eta} \frac{1}{t}
\frac{1}{ \Omega}
 [ \frac{ e^ {i ( pt + \Omega) R} - 1}{ pt + \Omega}  -
\frac{ e^ {i ( pt - \Omega) R} - 1}{ pt - \Omega}] ~.
\label{eq11}
\end{eqnarray}

\begin{figure}
\includegraphics[width=0.75\textwidth,clip=true]{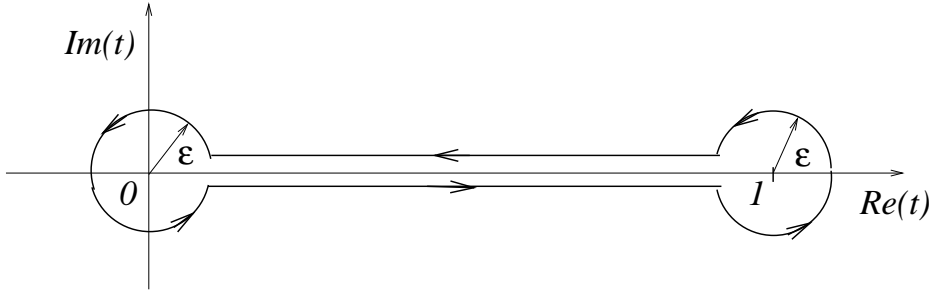}
\caption{The original path of integration $\Gamma$ used in Eq. (\ref{eq15}).}
\label{fig1}
\end{figure}

Here we like to distinguish the two cases $ p' > p $ and
$ p' < p$ and start with  $ p' > p $, where
 one can split (\ref{eq11}) as
\begin{eqnarray}
Y  =   \int_{ \Gamma} dt ( \frac{1-t}{t})^ { i \eta}
\frac{1}{t} \frac{1}{ \Omega}
 [ \frac{ e^ {i ( pt + \Omega) R}}{ pt + \Omega}  -
\frac{ e^ {i ( pt - \Omega) R}}{ pt - \Omega}]
 + 2 \int_{ \Gamma} dt ( \frac{1-t}{t})^ { i \eta} \frac{1}{t}
\frac{ 1}{ p^ 2 t^ 2 - \Omega^ 2}
\label{eq12}
\end{eqnarray}
since the poles of $\frac{ 1}{ p^ 2 t^ 2 - \Omega^ 2}$ do not
lie between $ t=0$ and $ t=1$. One has
\begin {eqnarray}
\frac{1}{ p^2 t^2 - \Omega^ 2}  =
  \frac{1}{ 2  \vec p \cdot \vec \Delta ( t  -
\frac{ \Delta^ 2}{ 2  \vec p \cdot \vec \Delta})} ~.
\label{eq13}
\end{eqnarray}
It is easily seen that for $ p' > p$  the pole position
  $ t_0 = \frac{ \Delta^ 2}{ 2  \vec p \cdot \vec \Delta}$
 as a function of $ \hat p \cdot \hat p~' $ is always larger in magnitude
than 1 if $ p > p' \hat p \cdot \hat p'$ and smaller than zero if
$ p < p' \hat p \cdot \hat p'$. Therefore
we can choose from the very beginning  the path $ \Gamma$ such that
the pole at $ t = t_0$
lies outside that closed path. Since the only singularities of the
integrand is the logarithmic cut
 between $t=0$ and $t=1$  and the pole at $t=0$  we choose the
$ \Gamma $ like in \cite{Gl09}.
For the convenience of the reader this is depicted in Fig.\ref{fig1}.

The second term in (\ref{eq12}) is easily evaluated.
If $ \Gamma$ is a circle with infinite radius the integral is zero. But
changing the path $ \Gamma$ to that circle one picks up a residue
due to (\ref{eq13}). If $ t_0 = \frac{ \Delta^
2}{ 2 \vec p \cdot \vec \Delta} > 1 $ then $ 1 - t_0 = | 1- t_0| e^ { i \pi} $
and $ t_0 = | t_0|$ ; if $ t_0 < 0$ then
$ 1 - t_0 = | 1- t_0| e^ { 2 i \pi} $ and $ t_0 = | t_0| e^ { i
\pi}$. In both cases $ ( \frac{ 1- t_0}{ t_0})^ { i \eta} =
( \frac{ |1- t_0|}{ |t_0|})^ { i \eta} e^ { -
\pi \eta}$. Consequently
\begin{eqnarray}
 2 \int_{ \Gamma} dt ( \frac{1-t}{t})^  { i \eta} \frac{1}{t}
\frac{1}{ p^2 t^2 - \Omega^ 2}
 =   - \frac{ 4 \pi i}{ \Delta^ 2}(
\frac{|  p^ 2  - {p'}^ 2|}{ \Delta^ 2}) ^ { i \eta}e^ { - \pi \eta} ~.
\label{eq14}
\end{eqnarray}

Now to the first term in (\ref{eq12}), denoted as $Y_1$.
 Again as in \cite{Gl09} we split the integral between
$t= \epsilon$ and $t= 1- \epsilon $ into two parts,
 choosing for instance $ t=1/2 $ as intermediate border, and
perform a partial integration for
 the integral $ \int_{ \epsilon}^ { 1/2}$ to
remove the pole $ \frac{1}{t} $. In this way we get
\begin{eqnarray}
Y_1 = \int_{zero} + \int_{ one} + ( 1 - e^ { - 2 \pi \eta})
( \int_{ \epsilon}^ {\frac{1}{2}} +
\int_{\frac{1}{2}}^ {1 - \epsilon})
\label{eq15}
\end{eqnarray}
and
\begin{eqnarray}
& &  \int_{ \epsilon}^ {\frac{1}{2}} dt
( \frac{1-t}{t})^ { i \eta} \frac{1}{t} \frac{1}{ \Omega}
 [ \frac{ e^ {i ( pt + \Omega) R} }{ pt + \Omega}  -
\frac{ e^ {i ( pt - \Omega) R}} { pt -\Omega}] = \cr
&& \frac{ -1}{ i \eta} ( \frac{1-t}{t})^ { i \eta}  \frac{1}{ \Omega}
 [ \frac{ e^ {i ( pt + \Omega) R} }{ pt + \Omega}  -
\frac{ e^ {i ( pt - \Omega) R} }
{ pt - \Omega}]|_{\epsilon}^ { \frac{1}{2}}\cr
& + &  \frac{ 1}{ i \eta}\int_{ \epsilon}^ {\frac{1}{2}} dt
t^ { - i \eta} \frac{d}{dt} ( 1 -t) ^ { i \eta}
\frac{1}{ \Omega}
  [ \frac{ e^ {i ( pt + \Omega) R} }{ pt + \Omega}  -
\frac{ e^ {i ( pt - \Omega) R} } { pt - \Omega}] ~.
\label{eq16}
\end{eqnarray}

The integral $\int_{zero}$  around $ t=0 $ is easily evaluated:
\begin{eqnarray}
 &   &   \int_{zero} dt ( \frac{1-t}{t})^ { i \eta} \frac{1}{t} \frac{1}{ \Omega}
 [ \frac{ e^ {i ( pt + \Omega) R} }{ pt + \Omega}  -
\frac{ e^ {i ( pt - \Omega) R} } { pt - \Omega}]\cr
& = &  \frac{2 i}{ \Delta^ 2}  cos \Delta R \epsilon^ { -i \eta}
\frac{1}{ \eta} ( 1 - e^ { - 2 \pi \eta}) ~.
\label{eq17}
\end{eqnarray}

It  cancels exactly against the lower limit contribution at
$ t= \epsilon $ in (\ref{eq16}).

The integral $\int_{ one}$  vanishes as $O( \epsilon) $ and the upper
limit
$( 1-\epsilon)$ in the last
integral in (\ref{eq15}) can be replaced by 1.
Thus as an intermediate result we have
\begin{eqnarray}
Y_1 & = &  ( 1 - e^ { - 2 \pi \eta})
(\frac{ -1}{ i \eta})  \frac{1}{ \Omega'}
 [ \frac{ e^ {i ( \frac{p}{2} + \Omega') R} }{ \frac{p}{2} + \Omega'}
 - \frac{ e^ {i (\frac{p}{2} - \Omega') R} } { \frac{p}{2} - \Omega'}]\cr
&  + &  (  1 - e^ { - 2 \pi \eta}) \frac{ 1}{ i \eta} \int_0^ {\frac{1}{2}} dt
 t^ { - i \eta} \frac{d}{dt} ( 1 -t ) ^ { i \eta} \frac{1}{ \Omega}
 [ \frac{ e^ {i ( pt + \Omega) R} }{ pt + \Omega}  -
\frac{ e^ {i ( pt - \Omega) R} } { pt - \Omega}]\cr
& +  & (  1 - e^ { - 2 \pi \eta}) \int_{\frac{1}{2}}^ 1 dt
( \frac{1-t}{t})^ { i \eta} \frac{1}{t}
 \frac{1}{ \Omega}
 [ \frac{ e^ {i ( pt + \Omega) R} }{ pt + \Omega}  -
\frac{ e^ {i ( pt - \Omega) R} }{ pt - \Omega}]
\label{eq18}
\end{eqnarray}
with
\begin{eqnarray}
\Omega' = \sqrt{ \frac{p^ 2}{4} - \vec p \cdot \Delta + \Delta^ 2} ~.
\label{eq19}
\end{eqnarray}

The lower integration limit $ \epsilon $ could be replaced by 0 since
only integrable logarithmic
singularities remain. The differentiation in t leads to several parts.
Only the pieces proportional to $R $ will survive in the
screening limit as will be argued below. They are given as
\begin{eqnarray}
 ( 1 - e^ { - 2 \pi \eta}) \frac{ R}{  \eta} \int_0 ^ \frac{1}{2}  dt
( \frac{ 1-t}{ t} ) ^ {  i \eta}
 \frac{1}{ \Omega} [ \frac{e^ { iR ( pt + \Omega)}}{ pt + \Omega}
( p + \frac{ d \Omega}{dt})
 -  \frac{ e^ { iR ( pt - \Omega)}}{pt -  \Omega}( p -
\frac{ d \Omega}{dt})] ~.
\label{eq20}
 \end{eqnarray}

The contribution from the lower limit $t=0$ is evaluated by the
method of steepest descent for $R \to \infty$. We use
\begin{eqnarray}
e^ { i R ( pt \pm \Omega)} & = &  e^ {\pm i R \Delta}
e^ { i R t( p \mp \frac{ \vec p \cdot \vec \Delta}{ \Delta})} ~, \\
p \pm \frac{ d \Omega}{dt}|_{ t=0} & = &  p  \mp
\frac{ \vec p \cdot \vec \Delta}{ \Delta} ~,
\end{eqnarray}
and obtain
\begin{eqnarray}
& &  ( 1 - e^ { - 2 \pi \eta}) \frac{ R}{  \eta} \int_0  dt
( \frac{ 1-t}{ t} ) ^ {  i \eta}
 \frac{1}{ \Omega} [ \frac{e^ { iR ( pt + \Omega)}}{ pt + \Omega}
( p + \frac{ d \Omega}{dt})
 -  \frac{ e^ { iR ( pt - \Omega)}}{pt -  \Omega}( p -  \frac{ d \Omega}{dt})]\cr
&  \to &   ( 1 - e^ { - 2 \pi \eta})\frac { i e^ { \frac{ \pi}{2}
     \eta}}
{ \eta \Delta^ 2} \Gamma( 1 - i
\eta) R^ { i \eta}
   [ e^ { i R \Delta} ( p - \frac{ \vec p \cdot
\vec \Delta}{ \Delta})^ { i \eta}
 + e^ { - i R \Delta} ( p + \frac{ \vec p \cdot
\vec \Delta}{ \Delta})^ { i \eta} ] ~.
\label{eq23}
\end{eqnarray}

Now all contributions related to the arbitrary $ t=1/2 $ border
should cancel each other. This is indeed
the case. By the same method  of steepest descent the asymptotic
contribution from the upper limit
 $ t=1/2 $ in (\ref{eq20}) can be
gained substituting $ t=1/2 -\tau $ and using
 \begin{eqnarray}
e^ { e^ { i R ( pt \pm  \Omega)}i} & = &   e^ { \pm i R ( \Omega' \pm p/2)}
e^ { - i R \tau ( p \mp  \frac{ d \Omega}{dt})} ~, \\
\frac{ d \Omega}{dt}|_{ t=1/2} & = &  \frac{-  \frac{p^ 2}{2}  +
\vec p \cdot \vec p'}{ \Omega'} ~,
\end{eqnarray}
as
\begin{eqnarray}
& &  ( 1 - e^ { - 2 \pi \eta}) \frac{ R}{  \eta} \int ^ \frac{1}{2}
dt ( \frac{ 1-t}{ t} ) ^ {  i \eta}
 \frac{1}{ \Omega} [ \frac{e^ { iR ( pt + \Omega)}}{ pt + \Omega}
( p + \frac{ d \Omega}{dt})
 -  \frac{ e^ { iR ( pt - \Omega)}}{pt -  \Omega}( p -  \frac{ d \Omega}{dt})]\cr
& \to &  \frac{ 1}{ i \eta \Omega'} ( 1 - e^ { - 2 \pi \eta})
e^ { i R \frac{p}{2}}
 ( \frac{ e^ { i \Omega' R}}{ p/2 + \Omega'} -
\frac{ e^ {-  i \Omega' R}}{ p/2 - \Omega'}) ~.
\label{eq26}
\end{eqnarray}
This cancels exactly against the first part in (\ref{eq18}).
Further the last integral in (\ref{eq18}) contributes at the lower limit
$ t= 1/2$ for $ R \to \infty$
\begin{eqnarray}
& &   (1 - e^ { - 2 \pi \eta}) \int_{\frac{1}{2}} dt
( \frac{1-t}{t})^ { i \eta} \frac{1}{t} \frac{1}{ \Omega}
 [ \frac{ e^ {i ( pt + \Omega) R} - 1}{ pt + \Omega}  -
\frac{ e^ {i ( pt - \Omega) R} - 1}
{ pt - \Omega}]\cr
& \to &  2 ( 1 - e^ { - 2 \pi \eta})  \frac{1}{\Omega'}
  [ \frac{ e^ { i R ( \Omega' + p/2)}} { \frac{p}{2} + \Omega'} \int_0 d\tau
 e^ { - i R \tau ( p +  \frac{ d \Omega}{dt})}
 -   \frac{ e^ { - i R ( \Omega' - p/2)}}{\frac{p}{2} - \Omega'}\int_0 d\tau
e^ { - i R \tau ( p -  \frac{ d \Omega}{dt})}] ~.
\label{eq27}
\end{eqnarray}

It is easily seen that for $ | \hat p \cdot \hat p' | \ne 1 $  the exponents
 $ p \pm \frac{ d \Omega}{dt}|_{ t= 1/2} \ne 0 $ and therefore that
limit is $ O(\frac{1}{ R})$
and can be neglected. By analogous steps one finds that
also the contribution from the upper limit $ t=1 $ of that last
integral
in (\ref{eq18}) is $O( \frac{1}{R})$ in
the screening limit. Therefore we obtain from (\ref{eq18}) and
 (\ref{eq23})
adding
the remaining parts of the
differentiation
\begin{eqnarray}
&&Y_1  \to   ( 1 - e^ { - 2 \pi \eta})\frac { i e^ { \frac{ \pi}{2}
    \eta}}
{ \eta \Delta^ 2}
\Gamma( 1 - i \eta) R^ { i \eta}
   [ e^ { i R \Delta} ( p - \frac{
\vec p \cdot \vec \Delta}{ \Delta})^ { i \eta}
 + e^ { - i R \Delta} ( p + \frac{ \vec p
\cdot \vec \Delta}{ \Delta})^ { i \eta} ]\cr
 &  + &    (  1 - e^ { - 2 \pi \eta}) \frac{ 1}{ i \eta}
\int_0^ {\frac{1}{2}} dt ( \frac{1-t}{t})^ {  i \eta}
 \frac{1}{ \Omega}\cr
[&(& - \frac{i \eta}{ 1 - t} -  \frac{1}{ \Omega} \frac{ d \Omega}{dt})
 [ \frac{ e^ {i ( pt + \Omega) R} }{ pt + \Omega}  -
\frac{ e^ {i ( pt - \Omega) R} } { pt - \Omega}]
 -    \frac{ e^ {i ( pt + \Omega) R}}{ (pt + \Omega)^ 2}
( p + \frac{ d \Omega}{dt})
  + \frac{ e^ {i ( pt - \Omega) R}} { (pt - \Omega)^ 2}
( p - \frac{ d \Omega}{dt})] ~.
\label{eq28}
\end{eqnarray}

Since there are no vanishing denominators nor $ pt \pm \Omega =0$
inside the range of integration the
remaining integral is $ O( \frac{1}{R})$ and one ends up with
the screening limit
\begin{eqnarray}
Y_1 & \to &  ( 1 - e^ { - 2 \pi \eta})
\frac { i e^ { \frac{ \pi}{2} \eta}} { \eta \Delta^ 2}
\Gamma( 1 - i \eta) R^ { i \eta}
   [ e^ { i R \Delta} ( p - \frac{ \vec p
\cdot \vec \Delta}{ \Delta})^ { i \eta}
 + e^ { - i R \Delta} ( p + \frac{ \vec p
\cdot \vec \Delta}{ \Delta})^ { i \eta} ] ~.
\label{eq29}
\end{eqnarray}
Then taking together with (\ref{eq14}) one finally  arrives  at
\begin{eqnarray}
Y & \to & ( 1 - e^ { - 2 \pi \eta})
\frac { i e^ { \frac{ \pi}{2} \eta}} { \eta \Delta^ 2}
\Gamma( 1 - i \eta) R^ { i \eta}
   [ e^ { i R \Delta} ( p -
\frac{ \vec p \cdot \vec \Delta}{ \Delta})^ { i \eta}
 + e^ { - i R \Delta} ( p +
\frac{ \vec p \cdot \vec \Delta}{ \Delta})^ { i \eta} ]\cr
&  - &  \frac{ 4 \pi i}{ \Delta^ 2}(
\frac{|  p^ 2  - {p'}^ 2|}{ \Delta^ 2}) ^ { i \eta}e^ { - \pi \eta} ~.
\label{eq30}
\end{eqnarray}

Thus like for the screening limit of the on-shell scattering
 amplitude
given in \cite{Gl09} there result two
terms, one,  as  expected, $R$-independent  and another still
dependent
on $ R $.

Now we turn to the case $ p' < p $ and start again from (\ref{eq11}), which
for a suitable path  $\Gamma$ can
again be brought into the form (\ref{eq12}). In this case the pole $ t_0 $
from (\ref{eq13})
lies on the real axis
between $ t=0 $ and $ t=1 $.

Since the path $ \Gamma$ encircles the cut the integral
\begin{eqnarray}
2 \int_{ \Gamma} dt ( \frac{1-t}{t})^ { i \eta} \frac{1}{t}
\frac{1} { p^ 2 t^ 2 - \Omega^ 2} =0
\label{eq31}
\end{eqnarray}
as is trivially seen by replacing the path $ \Gamma$ by a
circle with infinite radius.

\begin{figure}
\includegraphics[width=0.75\textwidth,clip=true]{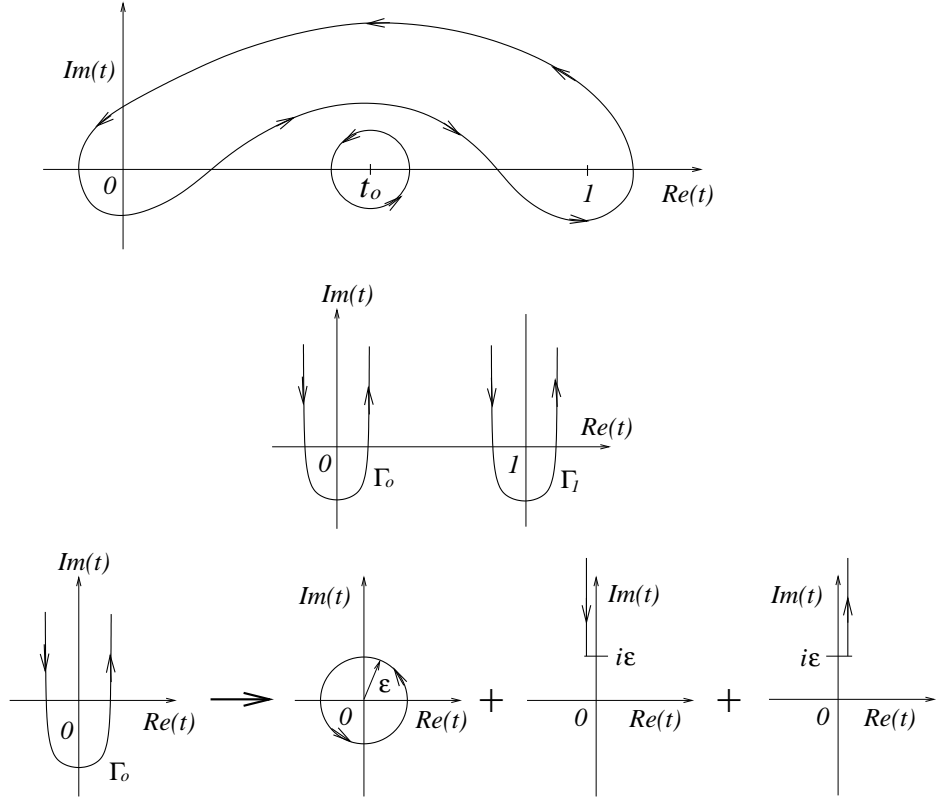}
\caption{The modifications of $\Gamma$ used in Eqs. (\ref{eq33})--(\ref{eq47}).}
\label{fig2}
\end{figure}

Thus we are left with the first term in (\ref{eq12}) which can be
brought into the form
\begin{eqnarray}
Y & = &   \int_{ \Gamma} dt
( \frac{1-t}{t})^ { i \eta} \frac{1}{t} \frac{1}{ \Omega}
 [ \frac{ e^ {i ( pt + \Omega) R} }{ pt + \Omega}  -
\frac{ e^ {i ( pt - \Omega) R} ( pt + \Omega) }
{ 2 \vec p \cdot \vec \Delta ( t -
\frac{ \Delta^ 2}{ 2 \vec p \cdot \vec \Delta})}] ~.
\label{eq32}
\end{eqnarray}
We deform the path $ \Gamma$ into $ \Gamma'$ such that the
lower part of $ \Gamma$ is moved
 between $ 0 $ and $  1 $ into the upper half plane, as shown in
Fig.\ref{fig2}.
 Thereby the path crosses the pole at
$ t_0 = \frac{ \Delta^ 2}{ 2 \vec p \cdot \vec \Delta}$,
 which leads to a residue:
\begin{eqnarray}
 \int_{ pole} dt ( \frac{1-t}{t})^ { i \eta} \frac{1}{t} \frac{1}{ \Omega} (- )
 \frac{ e^ {i ( pt - \Omega) R} ( pt + \Omega) }
{ 2 \vec p \cdot \vec \Delta ( t -
\frac{ \Delta^ 2}{ 2 \vec p \cdot \vec \Delta})}
 =    - \frac{ 4 \pi i}{ \Delta^ 2} ( \frac{ p^ 2 - {p'}^ 2}
{ \Delta^ 2})^ { i \eta} ~.
\label{eq33}
\end{eqnarray}
We used the fact that at the pole $ pt = \Omega$.

Next we further deform the path $ \Gamma'$  such that it
encircles $ t=0 $ coming from $ i \infty $
 and returning back to $ i \infty $ in the positive sense
and encircling $ t=1 $ again coming
 from $ i \infty $ and returning
back to $ i \infty $ in the positive sense.
This new paths $ \Gamma_0$ and $ \Gamma_1$  are displayed in
Fig.\ref{fig2}.

We get
\begin{eqnarray}
Y & = &  \int_{ \Gamma_0} dt
( \frac{1-t}{t})^ { i \eta} \frac{1}{t} \frac{1}{ \Omega}
 [ \frac{ e^ {i ( pt + \Omega) R} }{ pt + \Omega}  -
\frac{ e^ {i ( pt - \Omega) R} }{ pt - \Omega}]\cr
 &+&   \int_{ \Gamma_1} dt ( \frac{1-t}{t})^ { i \eta}
\frac{1}{t} \frac{1}{ \Omega}
 [ \frac{ e^ {i ( pt + \Omega) R} }{ pt + \Omega}  -
\frac{ e^ {i ( pt - \Omega) R} }{ pt - \Omega}] \cr
& - &  \frac{ 4 \pi i}{ \Delta^ 2} ( \frac{ p^ 2 - {p'}^ 2}{
\Delta^ 2})^ { i \eta} ~.
\label{eq34}
\end{eqnarray}

Because of the pole at $ t=0$ we separate the integral
$\int_{ \Gamma_0}$ into three parts (see Fig.\ref{fig2})
which read
\begin{eqnarray}
& & \int_{ \Gamma_0} dt ( \frac{1-t}{t})^ { i \eta} \frac{1}{t} \frac{1}{ \Omega}
 [ \frac{ e^ {i ( pt + \Omega) R} }{ pt + \Omega}  - \frac{ e^ {i ( pt
       - \Omega) R} }{ pt - \Omega}]\cr
& = &  \int_{ zero} dt \cdots +
( 1 - e^ { - 2 \pi \eta}) \int_{ i \epsilon}^ { i \infty} dtt^ { - i \eta-1}
  (1 -t)^  { i \eta}\frac{1}{ \Omega}
  [ \frac{ e^ {i ( pt + \Omega) R} }
{ pt + \Omega}  - \frac{ e^ {i ( pt - \Omega) R} }{ pt - \Omega}] ~.
\label{eq35}
\end{eqnarray}

The integral $\int_{ zero}$ is
\begin{eqnarray}
& & \int_{ zero} dt ( \frac{1-t}{t})^ { i \eta}
\frac{1}{t} \frac{1}{ \Omega}
 [ \frac{ e^ {i ( pt + \Omega) R} }{ pt + \Omega}  -
\frac{ e^ {i ( pt - \Omega) R} }{ pt - \Omega}]
 =  \int_{ - \frac{ 3 \pi}{2}}^ { \frac{ \pi}{2}} \epsilon i e^ { i \phi} d \phi
\frac{ e^ { - 2 \pi \eta}}{ ( \epsilon e^ { i \phi})^ { i \eta}}
 \frac{1}{ \epsilon  e^ { i \phi}}\frac{1}{ \Delta}
[ \frac{ e^ { i \Delta R}}{ \Delta} + \frac{ e^ {- i \Delta R}}{ \Delta}]\cr
& = &  e^ { - \frac{ 3 \pi}{2}\eta} ( 1 - e^ { - 2 \pi \eta}) \frac{ 2i}{ \Delta^ 2}
cos \Delta R \epsilon^ { -i \eta} \frac{1}{ \eta} ~.
\label{eq36}
\end{eqnarray}

Further
\begin{eqnarray}
& & ( 1 - e^ { - 2 \pi \eta}) \int_{ i \epsilon}^ { i \infty} dtt^ { - i \eta-1}
  (1 -t)^  { i \eta}\frac{1}{ \Omega}
 [ \frac{ e^ {i ( pt + \Omega) R} }{ pt + \Omega}  -
\frac{ e^ {i ( pt - \Omega) R} }{ pt - \Omega}]\cr
& = & ( 1 - e^ { - 2 \pi \eta}) [ - \frac{ 1}{ i \eta}
t^ { - i \eta} (1 -t)^  { i \eta}\frac{1}{ \Omega}
 [ \frac{ e^ {i ( pt + \Omega) R} }{ pt + \Omega}  -
\frac{ e^ {i ( pt - \Omega) R} }
{ pt - \Omega}]|_{ i \epsilon}^ { i \infty}\cr
& + & \frac{ 1}{ i \eta} \int_{ i \epsilon}^ { i \infty} dt
t^ { - i \eta} \frac{d}{dt} (1 -t)^  { i
\eta}\frac{1}{ \Omega}
 [ \frac{ e^ {i ( pt + \Omega) R} }{ pt + \Omega}  -
\frac{ e^ {i ( pt - \Omega) R} }{ pt - \Omega}]] ~.
\label{eq37}
\end{eqnarray}

There is no contribution at the integration limit $ i \infty$ and the
lower limit contribution cancels
against (\ref{eq36}).

Thus we are left with the intermediate result
\begin{eqnarray}
Y & = &  -   \frac{ 4 \pi i}{ \Delta^ 2} (
\frac{ p^ 2 - {p'}^ 2}{ \Delta^ 2})^ { i \eta}
 +  ( 1 - e^ { - 2 \pi \eta})
\frac{ 1}{ i \eta} \int_{ i \epsilon}^ { i \infty} dtt^ { - i \eta}
 \frac{d}{dt} (1 -t)^{ i \eta} \frac{1}{ \Omega}
 [ \frac{ e^ {i ( pt + \Omega) R} }{ pt + \Omega}  -
\frac{ e^ {i ( pt - \Omega) R} }{ pt - \Omega}] \cr
& +&  \int_{ \Gamma_1} dt ( \frac{1-t}{t})^ { i \eta}
\frac{1}{t} \frac{1}{ \Omega}
 [ \frac{ e^ {i ( pt + \Omega) R} }{ pt + \Omega}  -
\frac{ e^ {i ( pt - \Omega) R} }{ pt - \Omega}] ~.
\label{eq38}
\end{eqnarray}

The differentiation leads again to a piece explicitly proportional to $R$
\begin{eqnarray}
Y_R & \equiv &  ( 1 - e^ { - 2 \pi \eta}) \frac{R}{\eta}
\int_0^ { i \infty} dt ( \frac{1-t}{t})^ { i \eta}
\frac{1}{ \Omega}
 [ \frac{ e^ {i ( pt + \Omega) R} ( p + \frac{d \Omega}{dt}) }{ pt + \Omega}
 - \frac{ e^ {i ( pt - \Omega) R} ( p - \frac{d \Omega}{dt}) }{ pt -
   \Omega}] ~,
\label{eq39}
\end{eqnarray}
where we could put the lower limit of the integration to zero.
The integral converges at the upper limit
noting
\begin{eqnarray}
\Omega( i \tau)&  \to &    -  i \tau p  + \frac{\vec p
\cdot \vec \Delta}{  p}  ~, \\
\frac{d \Omega}{dt} & \to &  -p + \frac{\vec p \cdot \vec \Delta}{ i p
  \tau} ~.
\end{eqnarray}

Using
\begin{eqnarray}
e^ {i ( pt \pm \Omega) R} \to e^ { \pm i R \Delta}
e^ { - R \tau ( p \mp \frac{\vec p \cdot \vec \Delta}
{ \Delta})}
\label{eq42}
\end{eqnarray}
in the limit $ \tau \to 0$ we extract the leading behavior of $ Y_R $
\begin{eqnarray}
Y_R & \to & ( 1 - e^ { - 2 \pi \eta})\frac{iR}{\eta} \int_0 d \tau
( \frac{ e^ { 2 \pi i}} { i \tau})^ { i \eta}
 \frac{1}{ \Delta^ 2}
  [ e^ { i R \Delta} e^ { - R \tau ( p - \frac{\vec p \cdot \vec \Delta}{\Delta})}
( p - \frac{\vec p \cdot \vec \Delta}{\Delta}) \cr
 &+&   e^ {-  i R \Delta}  e^ { - R \tau ( p + \frac{\vec p
\cdot \vec \Delta}{\Delta})}
 ( p + \frac{\vec p \cdot \vec \Delta}{\Delta})]\cr
& = & ( 1 - e^ { - 2 \pi \eta})\frac{i}{\eta}e^ { - \frac{ 3 \pi
     \eta}{2}}
\frac{1}{ \Delta^ 2}
 [ e^ { i R \Delta} R ( p - \frac{\vec p \cdot \vec
        \Delta}{\Delta})
\int_0 d \tau \tau^ { - i \eta}
 e^ { - R \tau ( p - \frac{\vec p \cdot \vec \Delta}{ \Delta})}\cr
&  + &     e^ {-  i R \Delta} R  ( p + \frac{\vec p \cdot \vec
\Delta}{\Delta})\int_0 d \tau \tau^ { - i \eta}
 e^ { - R \tau ( p + \frac{\vec p \cdot \vec \Delta}{\Delta})}]\cr
& \to  &  (1 - e^ { - 2 \pi \eta})\frac{i}{\eta}
e^ { - \frac{ 3 \pi \eta}{2}} \frac{1}{ \Delta^ 2}
 R^ { i \eta} \Gamma( 1 - i \eta)
  [ e^ { i R \Delta}( p - \frac{\vec p \cdot \vec \Delta}{\Delta})^ { i \eta} +
e^ {-  i R \Delta}( p + \frac{\vec p \cdot \vec \Delta}{\Delta})^ { i
  \eta}] ~.
\label{eq43}
\end{eqnarray}

Next we regard the integral $ \int_{ \Gamma_1} $ from (\ref{eq38})
\begin{eqnarray}
& & \int_{ \Gamma_1} dt ( \frac{1-t}{t})^ { i \eta} \frac{1}{t} \frac{1}{ \Omega}
 [ \frac{ e^ {i ( pt + \Omega) R} }{ pt + \Omega}  -
\frac{ e^ {i ( pt - \Omega) R} }{ pt - \Omega}]\cr
& = & ( 1 - e^ { - 2 \pi \eta})\int_1 ^ { 1 + i \infty} dt
( \frac{1-t}{t})^ { i \eta} \frac{1}{t}
\frac{1}{ \Omega}
 [ \frac{ e^ {i ( pt + \Omega) R} }{ pt + \Omega}  -
\frac{ e^ {i ( pt - \Omega) R} }{ pt - \Omega}] ~.
\label{eq44}
\end{eqnarray}
To evaluate the leading contribution from the lower limit $ t=1 $ we need
\begin{eqnarray}
e ^ {i ( pt \pm \Omega) R} & \to &  e^ { i R( p \pm p')}
e^ {- R   \tau ( p \mp \frac{  \vec p \cdot \vec p'}{ p'})}
\label{eq45}
\end{eqnarray}
and obtain
\begin{eqnarray}
\int_{ \Gamma_1} dt \cdots &\to&  ( 1 - e^ { - 2 \pi \eta}) i
e^ { - \pi \eta} \frac{1}{p'}
  [ \frac{e^ { iR ( p + p')}}{ p + p'}  \int_0 d \tau
e^ { - R \tau (  p - \frac{  \vec p \cdot \vec p'}{ p'})}
 - \frac{e^ { iR ( p - p')}}{ p - p'}  \int_0 d \tau
e^ { - R \tau (  p + \frac{  \vec p \cdot \vec p'}{ p'})}] ~.
\label{eq46}
\end{eqnarray}
Since $ p \pm \frac{ \vec p \cdot \vec p'}{ p'} =
p( 1 \pm \hat p \cdot \hat p') \ne 0 $ for
 $ \theta \ne 0$, $\pi $, which we exclude,  the integrals in (\ref{eq46}) are
$ O(\frac{1}{R}) $.

Finally the additional terms resulting from the differentiation in
  (\ref{eq38})
are given as
\begin{eqnarray}
& & ( 1 - e^ { - 2 \pi \eta}) \frac{1}{ i \eta} \int_0^ { i \infty}
dt ( \frac{1-t}{t})^ { i \eta}
 \frac{ 1}{ \Omega}  \{ ( \frac{ - i \eta}{ 1-t} - \frac{1}{\Omega}
\frac{ d\Omega}{dt})\cr
& &  [ \frac{ e^ {i ( pt + \Omega) R} }{ pt + \Omega}  - \frac{
e^ {i ( pt - \Omega) R} }{ pt - \Omega}]
-    \frac{ e^ {i ( pt + \Omega) R} }{ (pt + \Omega)^ 2}
( p + \frac{ d\Omega}{dt})
 + \frac{ e^ {i ( pt - \Omega) R} }{ (pt - \Omega)^ 2} ( p - \frac{
   d\Omega}{dt})\} ~.
\label{eq47}
\end{eqnarray}
It can be shown that along the imaginary $t$-axis the
imaginary part of $ pt \pm \Omega $ is always
positive. Therefore that integral, too, vanishes like
$ O( \frac{1}{R}) $ in the screening limit.

Thus we are finally left for $ p' <p $ with
\begin{eqnarray}
Y & \to & -  \frac{ 4 \pi i}{ \Delta^ 2} ( \frac{ p^ 2 - {p'}^ 2}
{ \Delta^ 2})^ { i \eta}\cr
& + & (1 - e^ { - 2 \pi \eta})\frac{i}{\eta}
e^ { - \frac{ 3 \pi \eta}{2}} \frac{1}{ \Delta^ 2}
 R^ { i \eta} \Gamma( 1 - i \eta)
  [ e^ { i R \Delta}( p - \frac{\vec p \cdot \vec \Delta}{\Delta})^ { i \eta} +
e^ {-  i R \Delta}( p + \frac{\vec p \cdot \vec \Delta}{\Delta})^ { i
  \eta}] ~.
\label{eq48}
\end{eqnarray}
This is now ´to be compared with (\ref{eq30}), repeated for the
convenience of the reader and valid for $ p' > p$
\begin{eqnarray}
Y & \to & -   \frac{ 4 \pi i}{ \Delta^ 2}( \frac{ {p'}^ 2 -   p^ 2 }
{ \Delta^ 2}) ^ { i \eta} e^ { - \pi \eta}\cr
& + & (1 - e^ { - 2 \pi \eta})\frac{i}{\eta} e^ { \frac{ \pi}{2} \eta}
\frac { 1 } {  \Delta^ 2}
R^ { i \eta} \Gamma( 1 - i \eta)
  [ e^ { i R \Delta} ( p - \frac{ \vec p \cdot
\vec \Delta}{ \Delta})^ { i \eta}
 + e^ { - i R \Delta} ( p + \frac{ \vec p
\cdot \vec \Delta}{ \Delta})^ { i \eta} ] ~.
\label{eq49}
\end{eqnarray}

We see that the $ R$-independent part jumps
from $ p' < p $ to $ p' > p $ by a factor $ e^ { - \pi \eta} $
whereas the oscillating $ R $-dependent part
jumps by a factor $ e^ { 2 \pi \eta} $.

Lastly we turn to the half-shell t-matrix given in (\ref{eq10}). The
asymptotic value of $ A $ as derived in
\cite{Gl09}
is
\begin{eqnarray}
A \to \frac{ 1} { ( 2\pi)^ { \frac{3}{2}}} e^ { - \frac{\pi}{2} \eta }
( 2pR)^ { - i \eta} \Gamma( 1 + i \eta)
\label{eq50}
\end{eqnarray}
and therefore the prefactor in (\ref{eq10}) using (\ref{eq6}) is asymptotically
\begin{eqnarray}
- 2 \pi \frac{e^ 2}{ ( 2 \pi)^ { \frac{3}{2}}} A C( - i \eta,1) \to
i \frac{e^ 2}{ ( 2 \pi)^3}
 e^ { \frac{\pi}{2}\eta} ( 2pR)^ { - i \eta} \Gamma( 1 + i \eta) ~.
\label{eq51}
\end{eqnarray}

This leads for $ p' < p $ to the half shell t-matrix element in
the screening limit
\begin{eqnarray}
& & < \vec p~' | V_R | \Psi_R^ { (+)} > \to  \frac{e^ 2}{ 2 \pi^ 2}
 e^ { \frac{\pi}{2}\eta} ( 2pR)^ { - i \eta} \Gamma( 1 + i \eta)
\frac{1}{ \Delta^ 2}
 ( \frac{ p^ 2 - {p'}^ 2}{ \Delta^ 2})^ { i \eta}\cr
& - &  \frac{e^ 2}{ ( 2 \pi)^2}  \frac{1}{ \Delta^ 2}
 [ e^ { i R \Delta}( \frac{1}{2} (p - \frac{\vec p \cdot
\vec \Delta}{\Delta}))^ { i \eta} +
e^ {-  i R \Delta}(\frac{1}{2}( p + \frac{\vec p \cdot
\vec \Delta}{\Delta}))^ { i \eta}] ~,
\label{eq52}
 \end{eqnarray}
where we used $ \Gamma(1 + i \eta) \Gamma( 1 - i \eta) =
\frac{ \pi \eta}{ sinh \pi \eta}$.

On the other hand the pure half shell t-matrix is well known
\cite{Kok} and given for $ p' < p $ by
\begin{eqnarray}
 < \vec p~' | V_C | \Psi_{\vec p}^ {C (+)} > = \frac{e^ 2}{ 2 \pi^ 2}
 e^ { \frac{\pi}{2}\eta} \Gamma( 1 + i \eta)   \frac{1}{ \Delta^ 2}
 ( \frac{ p^ 2 - {p'}^ 2}{ \Delta^ 2})^ { i \eta} ~.
\label{eq53}
\end{eqnarray}

Therefore  for $ p' < p $  we find the following  result in the
screening limit
\begin{eqnarray}
 & & < \vec p~' | V_R | \Psi_R^ { (+)} > \to  e^ { - i \eta ln  2pR }
 < \vec p~' | V_C | \Psi_{\vec p}^ {C (+)} >\cr
& - & \frac{e^ 2}{ ( 2 \pi)^2}  \frac{1}{ \Delta^ 2}
 [ e^ { i R \Delta}( \frac{1}{2} (p - \frac{\vec p
\cdot \vec \Delta}{\Delta}))^ { i \eta} +
e^ {-  i R \Delta}(\frac{1}{2}( p + \frac{\vec p
\cdot \vec \Delta}{\Delta}))^ { i \eta}] ~.
\label{eq54}
 \end{eqnarray}

The first term is the expected one as given in~\cite{Ford64}. But there
is, like for the on-shell t-matrix, an
additional term, which only after integration over some angular
region would disappear in the screening
limit.

In case of $  p' > p $ the pure half shell t-matrix differs
by a factor $ e^ { - \pi \eta} $ and is
\begin{eqnarray}
 < \vec p~' | V_C | \Psi_{\vec p}^ {C (+)} > = \frac{e^ 2}{ 2 \pi^ 2}
 e^ { -\frac{\pi}{2}\eta} \Gamma( 1 + i \eta)   \frac{1}{ \Delta^ 2}
 ( \frac{ {p'}^ 2 - p^ 2}{ \Delta^ 2})^ { i \eta} ~.
\label{eq55}
\end{eqnarray}

Therefore in this case and  using (\ref{eq49}) we find the following
result in the screening limit
\begin{eqnarray}
 & & < \vec p~' | V_R | \Psi_R^ { (+)} > \to  e^ { - i \eta ln  2pR }
 < \vec p~' | V_C | \Psi_{\vec p}^ {C (+)} >\cr
& - & e^ { 2 \pi \eta} \frac{e^ 2}{ ( 2 \pi)^2}  \frac{1}{ \Delta^ 2}
 [ e^ { i R \Delta}( \frac{1}{2} (p - \frac{\vec p
\cdot \vec \Delta}{\Delta}))^ { i \eta} +
e^ {-  i R \Delta}(\frac{1}{2}( p + \frac{\vec p
\cdot \vec \Delta}{\Delta}))^ { i \eta}] ~.
\label{eq56}
\end{eqnarray}

The first term has the same structure as above, but the
second one differs by the factor
 $ e^ { 2 \pi\eta} $ from the one above.

The first term has the same structure as above, but the
second one differs by the factor
 $ e^ { 2 \pi\eta} $ from the one above.

\section{Numerical results}
\label{numresults}

It is interesting to compare the derived asymptotic
forms (\ref{eq54}) and (\ref{eq56})
to the numerical solutions of the Lippmann-Schwinger
equation for the sharply cut off Coulomb potential with different
cut-off radii.
As is well known \cite{elster1998,Gl09}
this equation can be written as a
two-dimensional integral equation
\begin{eqnarray}
 T( q^\prime , q, x^\prime) = \frac1{2 \pi} v( q^\prime , q, x^\prime, 1 )
+\int\limits_0^\infty d q''  {q''}^{\, 2}
 \int\limits_{-1}^1 d x'' v( q^\prime , q'' , x^\prime, x'' )  \frac1{ z -
\frac{{q''}^{\, 2}}{m}}
 T( q'' , q, x'' ) ,
\label{2dimLSE}
\end{eqnarray}
where \begin{eqnarray}
v( q^\prime , q , x^\prime, x )  =
 \int\limits_{0}^{2 \pi} d \varphi V( q^\prime, q, x^\prime x + \sqrt{ 1 -
{x^\prime}^{\, 2} } \sqrt{ 1 - {x}^{\, 2} } \cos \varphi ) \label{smallv}
\end{eqnarray}
and $m$ is the reduced mass of the system.

For the sharply screened Coulomb potential of the range $R$ considered in
this paper \begin{eqnarray}
V ( q^\prime, q , y ) = \frac{e^2}{ 2 \pi^2 } \,
               \frac{ 1 - \cos ( Q R ) }{ Q^2 },
\end{eqnarray}
where $ Q \equiv \sqrt{ {q^\prime}^{\, 2} + q^2 - 2 q^\prime \, q y }$
and the integral over $\varphi$ in Eq.~(\ref{smallv}) is carried out numerically.

\begin{figure}
\includegraphics[width=0.75\textwidth,clip=true]{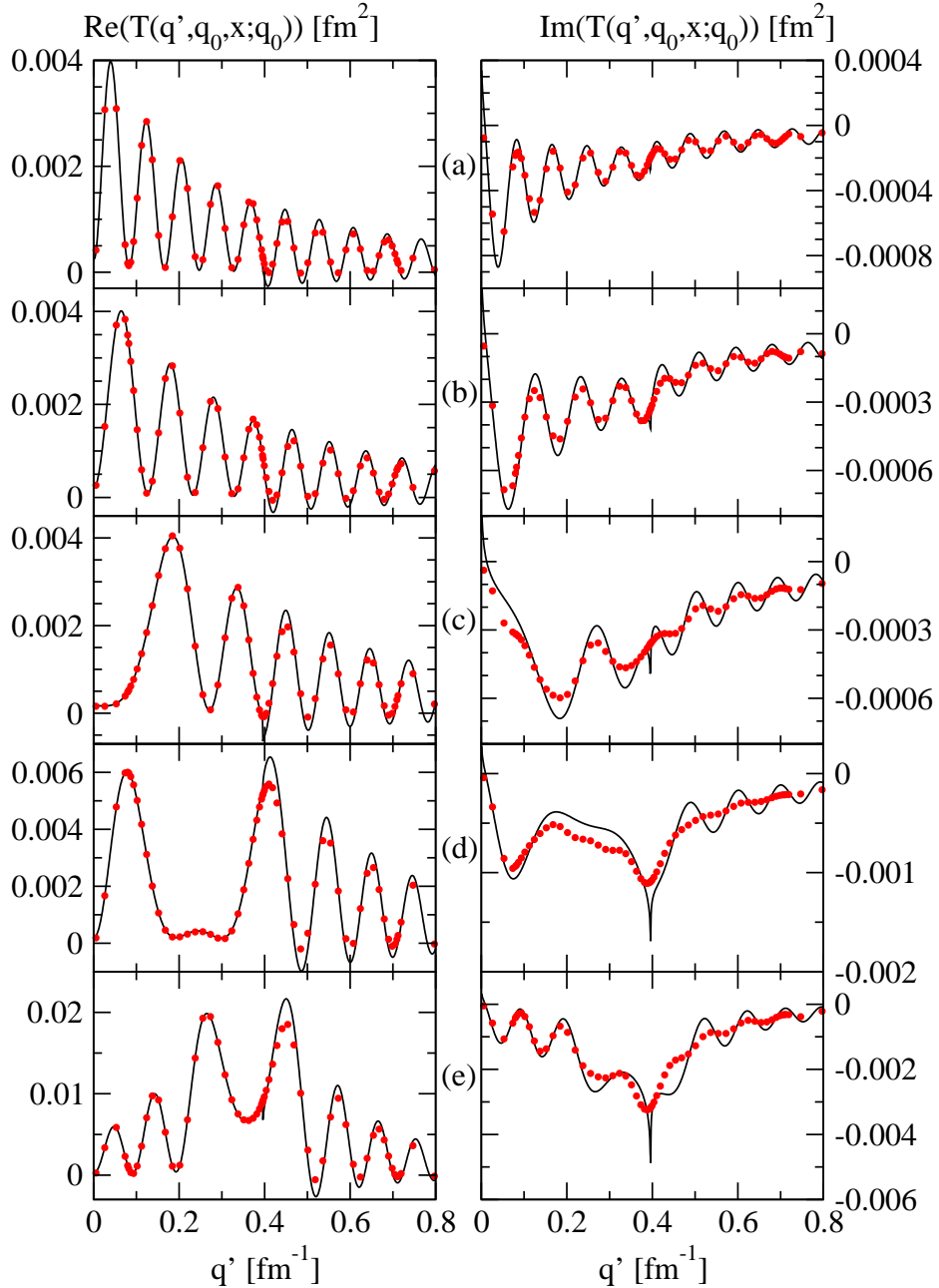}
\caption{(color online) The real (left) and imaginary (right)
part of the half-shell t-matrix $T( q^\prime , q_0, x ; \, q_0)$
for the sharply cut-off
Coulomb potential
 with the cut-off  radius $R=80$~fm.
The solid (black) line represents the asymptotic analytical expression
given in (\ref{eq54}) and (\ref{eq56}).
The (red) dots show our numerical results. From top to bottom five different
values of $x$ are chosen:
(a) $x=-0.91$,
(b) $x=-0.50$,
(c) $x= 0.03$,
(d) $x= 0.62$,
(e) $x= 0.90$.
}
\label{fig3}
\end{figure}

\begin{figure}
\includegraphics[width=0.75\textwidth,clip=true]{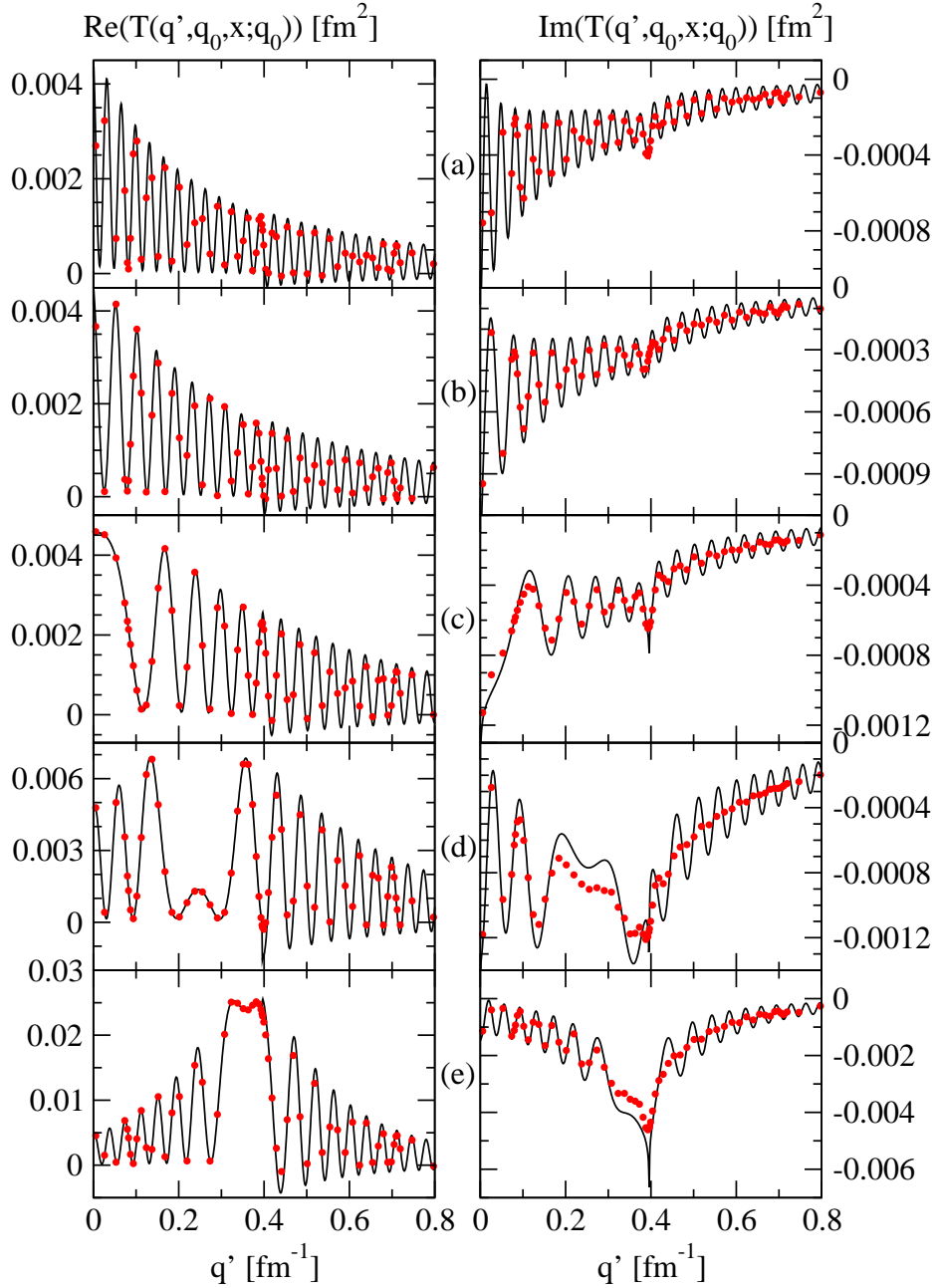}
\caption{(color online) The same as in Fig.~3 but with the cut-off radius $R=200$~fm.}
\label{fig4}
\end{figure}

Solving the two-dimensional equation (\ref{2dimLSE}) is a difficult
numerical problem because $ V ( q^\prime, q , x ) $ shows a highly oscillatory
behavior, especially for large $R$.
We solved (\ref{2dimLSE})
for positive energies where
\begin{eqnarray}
z = E_{c.m.} + i \epsilon \equiv \frac{q_0^2}{m} + i \epsilon
\end{eqnarray}
by generating the corresponding Neumann series and
summing it up by Pad\`e. In each iteration the Cauchy singularity
was split into a principal-value
integral (treated by subtraction) and a $\delta$-function piece.
All details about our numerical performance are given in \cite{Gl09}.
By solving (\ref{2dimLSE}) we obtain all
matrix elements $  T( q^\prime , q, x ; \, q_0)$; they
can be chosen on-shell (as investigated in \cite{Gl09}), half-shell
or totally off-shell. Here we are interested in the half-shell
elements, $  T( q^\prime , q_0, x ; \, q_0)$,
and show examples
in Figs.~\ref{fig3}--\ref{fig4}.
We choose just five (more or less arbitrary) values of $x=$-0.91,
-0.50, 0.03, 0.62 and 0.90, which corresponds
to the following angles $\theta$ between vectors
${\vec q}^{\, \prime} $ and ${\vec q}_0$:
$\theta=$155.5$^\circ$,
120.0$^\circ$,
88.3$^\circ$,
51.7$^\circ$ and
25.8$^\circ$.
Then for each fixed value of $x$ we display the half-shell
matrix elements $  T( q^\prime , q_0, x ; \, q_0)$ as
a function of $ q^\prime $.
The analytical asymptotic forms of Eqs.~(\ref{eq54}) and (\ref{eq56})
are given by the line and the numerical results are shown with symbols.
We concentrate on the $ q^\prime $ region in the vicinity
of $q_0$ where the most interesting structures appear
and skip the region of higher $ q^\prime $ values, where
$  T( q^\prime , q_0, x ; \, q_0)$ tends to zero showing more or less
rapid oscillations.
As in \cite{Gl09} we restrict ourselves to the system of two protons
scattering solely by the Coulomb force
at $E_p^{lab}=13$~MeV. This gives $q_0 \approx 0.396$~fm$^{-1}$.
Two cases of the cut-off radii
$R=80$~fm (Fig.~\ref{fig3}) and
$R=200$~fm (Fig.~\ref{fig4}) are considered.

In the case of $R=80$~fm the real part of
$T( q^\prime , q_0, x ; \, q_0)$ is usually by one order
of magnitude bigger than the imaginary part. The exception
is $x= 0.62$, where the both parts are comparable.
The analytical asymptotic form agrees rather well with the numerical
result for the real part. The agreement is in fact very good
in the region of $q^\prime < q_0$ and a bit less satisfactory
for $q^\prime > q_0$. For the imaginary part there are clear deviations
between the analytical and numerical results, which become more pronounced
for $x \ge 0$. In particular the analytical results show much more
oscillatory behavior for $q^\prime > q_0$. Note also a sharp structure around
$q^\prime = q_0$ which develops for the imaginary part at $x \ge 0$.

For $R=200$~fm the real part of
$T( q^\prime , q_0, x ; \, q_0)$ is clearly dominant for all the
considered values of $x$.
The analytical asymptotic form shows much more oscillations
than for $R=80$~fm. It is clear that in order to trace these oscillations,
many more points in the numerical solution would be required.
Despite this fact, one can see at least fair agreement between
the asymptotic analytical results and numerical solutions
at the calculated points in the case of the real part.
For the imaginary part, like in the case of $R=80$~fm,
the agreement is worse. This is presumably caused by limitations
of our numerical treatment.

\section{Summary and conclusions}
\label{summary}

We investigated the screening limit of the
exact analytical three-dimensional half-shell
t-matrix for a sharply cut-off Coulomb potential.
We used the exact three-dimensional wave
function for a sharply cut-off Coulomb potential derived in \cite{Gl09}.
Our direct three-dimensional approach avoids problems
related to the summation of the infinite
number of partial wave components. Numerical solutions of the
three-dimensional Lippmann-Schwinger equation for large cut-off radii
agree fairly well with the asymptotic values.

\section*{Acknowledgments}
This work was supported by the 2008-2011 Polish Science Funds as a
research project No. N N202 077435. It was also partially supported by the
Helmholtz
Association through funds provided to the virtual institute ``Spin
and strong QCD''(VH-VI-231) and by
  the European Community-Research Infrastructure
Integrating Activity
``Study of Strongly Interacting Matter'' (acronym HadronPhysics2, Grant
Agreement n. 227431)
under the Seventh Framework Programme of EU.
 The numerical calculations were
performed on the IBM Regatta p690+ of the NIC in J\"ulich,
Germany.

\end{document}